\documentclass{sf2a-conf2018}
\usepackage{graphicx}
\usepackage{hyperref}
\usepackage[]{natbib}  
\usepackage{epstopdf}

\def\BibTeX{{\rm B\kern-.05em{\sc i\kern-.025em b}\kern-.08em
    T\kern-.1667em\lower.7ex\hbox{E}\kern-.125emX}}
\bibpunct{(}{)}{;}{a}{}{,}  

\begin{document}

\TitreGlobal{SF2A 2018}


\title{The BeSS database: a fruitful professional-amateur collaboration}

\runningtitle{The BeSS database}

\author{C. Neiner}\address{LESIA, Paris Observatory, PSL University, CNRS, Sorbonne Université, Univ. Paris Diderot, Sorbonne Paris Cité, 5 place Jules Janssen, 92195 Meudon, France}

\setcounter{page}{237}


\maketitle


\begin{abstract}
BeSS is a database containing a catalogue of Be stars and their spectra, set up more than 10 years ago as a collaboration between professional and amateur astronomers. It currently contains over 177000 spectra of 2340 stars, provided by $\sim$150 different observers. Its continuous success has already led to the use of BeSS data in more than 70 scientific papers. 
\end{abstract}



\section{Be stars}
\subsection{Classical Be stars}

Classical Be stars are non-supergiant hot stars (typically O7 to A2) that show or have shown emission lines in their spectrum \citep[see a complete review by][]{rivinius2013}. This concerns about 20\% of all B stars. The emission comes from a circumstellar disk fed by discrete ejections of matter from the star, called outbursts. Classical Be stars are known to be very rapid rotators (of the order of 250 km/s). As a consequence these stars are flattened by the centrifugal force and their line profiles are broad due to the Doppler effect. In addition, classical Be stars host pulsations. The combination of rapid rotation and pulsations is thought to be at the origin of the outbursts \citep{neiner2014}.

Thanks to the many physical processes at work in classical Be stars, these stars are great laboratories to study stellar physics. However, due to these many processes, they vary photometrically and spectroscopically on all timescales. Therefore the best way to study them and their physical processes is to observe them very regularly to investigate their variations in detail.

For example, the shape of the emission lines depends on the inclination under which the disk is observed \citep{slettebak1992}. Moreover, an asymmetry between the blue and red sides of the emission lines can be observed if the disk is not homogeneous \citep{telting1994}. Since the disk is fed by sporadic ejections of matter from the star and slowly leaks into the interstellar medium, the level of emission also changes with time \citep[e.g.][]{koubsky2000}. In addition, the photospheric lines vary because of the pulsations and possible surface spots. 

\subsection{Herbig Be stars}

Herbig Be stars are pre-main sequence B stars, i.e. that are still in their formation stage and not burning hydrogen in their core yet \citep[see the review by][]{alecian2011}. Thus they still host an accretion disk, which produces emission in their spectra, similar to the one observed in classical Be star spectra. The spectrum of both types of objects can however be easily differentiated in the infrared, where Herbig Be stars show an excess of light due to dust while classical Be stars show an infrared excess due to their hydrogen disk. In addition, about 10\% of Herbig Be stars are magnetic \citep{alecian2013} and therefore they may also display magnetospheric emission.

The line emission of Herbig Be stars varies as accretion events occur. Studying the shor-term variation of this emission thus provides a direct signature of accretion and will help to understand the formation processes of hot stars.

\subsection{B[e] supergiants}

B[e] stars regroup several types of objects that display emission in their spectrum including in forbidden lines. IR dust emission is also observed, as in Herbig Be stars. This very diverse group includes, e.g., compact planetary nebulae and symbiotic stars. Here, we consider only B[e] supergiants, i.e. post-main sequence luminous stars \citep[see][]{kraus2016}. They host hot fast winds and a circumstellar gas+dust disk at the origin of their spectral emission. 

With their release of huge amounts of energy and material in their environment, B[e] supergiants play a crucial role in the evolution of their host galaxies. However, B[e] supergiants are currently not predicted by stellar evolution models and their exact evolutionary state is thus not known. The physical mechanism causing the formation of their dusty disk is also unclear. Binarity may however play a role. Regular observations of the emission variation may provide new clues about these stars \citep[e.g.][]{maravelias2017}.

\subsection{Magnetic Be stars}

Magnetic Be stars are O, B or A stars that host a magnetic field \citep[see the review by][]{grunhut2015}. This concerns about 10\% of all hot stars. Due to the presence of a, usually dipolar, magnetic field, the wind particles are channeled along the field lines around the star. Particles coming from both magnetic poles collide in the magnetic equatorial plane. Depending on the relative values of the wind velocity, field strength and stellar rotation, i.e. on whether the Keplerian co-rotation radius is larger or smaller than the Alfven radius, this material can remain trapped in a centrifugally-supported magnetosphere. Otherwise, the material falls back onto the star. In this case, a magnetosphere still exists but is then a dynamical magnetosphere since its content is constantly renewed. Both types of magnetospheres can become dense enough to produce emission visible in the spectra of magnetic Be stars \citep[see][for more details]{owocki2014}.

Since the magnetic axis is usually not aligned with the rotation axis of the star, the magnetosphere is viewed under various inclinations as the star rotates. This rotational modulation gives rise to periodic variations of the emission line profiles. This periodicity allows to easily distinguish magnetic Be stars from other Be stars, even without a magnetic measurement.

\subsection{$\gamma$\,Cas analogs} 

$\gamma$\,Cas was the first Be star discovered \citep{secchi1867} and for a long time considered as the prototype of classical Be stars. It turned out later not to be a classical Be star, as it displays a very peculiar X-ray spectrum, with anomalously luminous, hard, and variable X-ray flux \citep[see the review by][]{smith2016}. Only $\sim$20 other examples of such Be stars have been found until now \citep{naze2018} and are called $\gamma$\,Cas analogs.

The origin of the X-ray emission of $\gamma$\,Cas analogs is not understood at all. Accretion disk scenarios have been proposed, as well as magnetic interaction between the star and a decretion disk. Observations of the emission spectra of $\gamma$\,Cas analogs may allow us to finally understand these objects. 

\section{The BeSS database}

Following the many variations of classical Be stars requires regular observations of their spectra. This is why the BeSS database\footnote{http://basebe.obspm.fr} has been set up in 2007 \citep{neiner2011}. Subsequently, other types of Be stars have also been added to BeSS. 

\begin{figure}[t!]
 \centering
 \includegraphics[width=0.65\textwidth,clip]{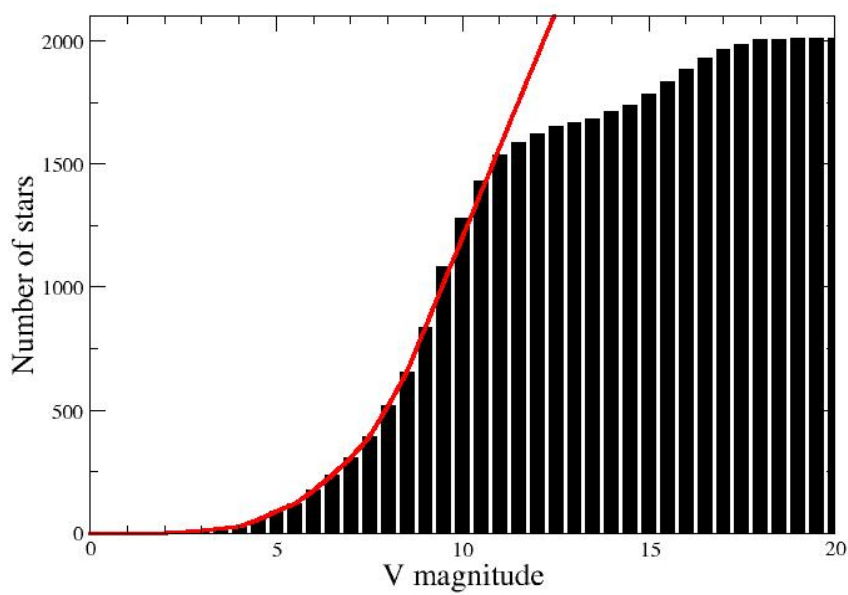}      
  \caption{Cumulative number of stars in the BeSS catalogue with respect to the V magnitude. The catalogue is complete below V$\sim$11, while many Be stars remain to be discovered above that limit.}
  \label{BeSS_mag}
\end{figure}

As of today, it contains a catalogue of all known classical Be stars, all known Herbig Be stars, all known B[e] supergiants, as well as some Oe stars that have emission due to their wind and $\gamma$\,Cas analogs. In the future we plan to also add magnetic Be stars. The catalogue is complete up to magnitude V$\sim$11 (see Fig.~\ref{BeSS_mag}). Above this magnitude, many Be stars remain to be discovered. For each star, the catalogue gathers information on its parameters, such as coordinates, magnitude, temperature, gravity, projected rotation velocity, spectral type, etc.

In addition to the catalogue of stars, the BeSS database host spectra of these objects provided by anyone willing to share their data. At the time of writing, the database contains 177590 spectra, including $\sim$60\% provided by amateur astronomers. Each spectrum uploaded in BeSS first goes through an automatic validation: check of its header keywords and plausibility checks (e.g., can that star indeed be observed at that time from that location?). Then a human perfoms final checks directly on the spectrum. All amateur spectra are validated by amateur astronomers trained for that purpose. Specific types of stars, e.g. B[e] supergiants, are validated by a professional astronomer specialized in this type of targets. 

\section{Contribution of amateurs to BeSS}

Amateurs (and professional) astronomers can contribute to BeSS in many ways.

First, one can acquire H$\alpha$ spectra of Be stars to participate in the regular survey of all Be star disks. It is best to use high-resolution spectrographs (R$>$6000) whenever possible and keep low-resolution spectrographs only for faint targets. When an oberver notices a sudden increase in emission, meaning an outburst has started, he/she alerts the community so that others can acquire additional spectra in the following days and weeks. Such follow-up campaign are very useful to fully characterize the matter ejection. Quick alerts can be posted on the spectro-l mailing list\footnote{spectro-l@yahoogroups.com} and those targets are also identified in ArasBeAm (see below).

Moreover, spectra obtained at other wavelengths than H$\alpha$ are also welcome in BeSS, e.g. at H$\beta$, He{\sc i} lines, in the UV or in the infrared, as they provide additional information on the star or on its disk.

Campaigns are also organised in conjunction with other instruments or satellites. Recent examples include providing spectra of the Be targets of the BRITE constellation of nanosatellite, which photometrically observes bright stars, and surveying active Be stars to give the alert for interferometric observations of the onset of outbursts with CHARA. Such observing programs are usually set up by a professional astronomer but many amateurs contribute to the data collection.

Finally, observers can also perform spectroscopy of B stars to discover new faint Be stars to be added to the BeSS catalogue. Usually, B stars are first observed with low-resolution spectrographs to search for signs of emission and then confirmed as Be stars with a high-resolution spectrum. 21 new Be stars have already been identified this way by amateur astronomers. 

Those amateurs who do not host a spectroscopic equipement can still participate in BeSS, either by using spectrographs available in large observatories such as at the T60 telescope at the Pic du Midi (France), at Saint V\'eran Observatory (France) or at universities over the world, or by participating in the data analysis of BeSS, e.g. to measure equivalent width variations of spectral lines, variations in the emission profiles, or to determine stellar parameters from the spectra. 

Another important contribution of amateurs to BeSS is the ArasBeAm tool\footnote{http://www.arasbeam.fr}. It allows anyone to know which Be stars to observe from their location that day. Not only does it display the targets that are observable, but it shows a coloured-coded priority ranking depending on whether observations of those targets are needed in BeSS or not. This priority depends on the level of activity of the star, on ongoing follow-up campaigns, and on spectra already recently obtained by others. 

Finally, amateurs can contribute or even lead publications resulting from the BeSS data. At the time of writing, 71 scientific publications have made use of BeSS data, including 6 with an amateur astronomer as the first author. Amateurs also publish the BeSS monthly report, which summarizes all the new data and findings from the past month in BeSS. 

\section{The success of BeSS}

The success of BeSS lies on several reasons. First, Be stars are very interesting objects for amateurs to observe as their spectrum keeps changing. Moreover, the ArasBeAm tool gives them a strong incentive to observe those targets knowing that the data will be useful to science. The number of published papers also shows that the data are indeed being used. In addition, the human validation of the spectra makes both the database a reliable source of good spectra for astronomers and a good way for the beginners in spectroscopy to get feedback on their data and improve their observing technique. Therefore both uploaders and downloaders of data find an interest in using BeSS. Finally, spectrographs (e.g. by Shelyak Instruments\footnote{https://www.shelyak.com/}) and softwares (e.g. VisualSpec\footnote{http://astrosurf.com/vdesnoux/} or ISIS\footnote{http://www.astrosurf.com/buil/isis-software.html}) have been developed for amateurs by amateurs and allow to easily produce spectra to the BeSS quality  standard and in the BeSS fits format. 

\begin{acknowledgements}
This work as made used of the SIMBAD database operated at CDS, Strasbourg (France), of NASA's Astrophysics Data System (ADS), and of the BeSS database, operated at LESIA, Observatoire de Meudon, France: http://basebe.obspm.fr.

I would like to thank the BeSS validators (C. Buil, F. Cochard, V. Desnoux, O. Garde, T. Garrel, M. Kraus, T. Lemoult, and O. Thizy) for their huge work and dedication in keeping the validation time short, the administrators of BeSS and ArasBeAm (B. de Batz, F. Cochard, J.-J. Broussat),  V. Desnoux for preparing the BeSS Monthly report, and all the BeSS observers for their contribution to the database.
\end{acknowledgements}

\bibliographystyle{aa}  
\bibliography{sf2a-Neiner-BeSS} 

\begin{thebibliography}{15}
\expandafter\ifx\csname natexlab\endcsname\relax\def\natexlab#1{#1}\fi

\bibitem[{{Alecian}(2011)}]{alecian2011}
{Alecian}, E. 2011, in IAU Symposium, Vol. 272, Active OB Stars: Structure,
  Evolution, Mass Loss, and Critical Limits, ed. C.~{Neiner}, G.~{Wade},
  G.~{Meynet}, \& G.~{Peters}, 354--365

\bibitem[{{Alecian} {et~al.}(2013){Alecian}, {Wade}, {Catala}, {Grunhut},
  {Landstreet}, {Bagnulo}, {B{\"o}hm}, {Folsom}, {Marsden}, \&
  {Waite}}]{alecian2013}
{Alecian}, E., {Wade}, G.~A., {Catala}, C., {et~al.} 2013, \mnras, 429, 1001

\bibitem[{{Grunhut} \& {Neiner}(2015)}]{grunhut2015}
{Grunhut}, J.~H. \& {Neiner}, C. 2015, in IAU Symposium, Vol. 305, Polarimetry,
  ed. K.~N. {Nagendra}, S.~{Bagnulo}, R.~{Centeno}, \& M.~{Jes{\'u}s
  Mart{\'{\i}}nez Gonz{\'a}lez}, 53--60

\bibitem[{{Koubsk{\'y}} {et~al.}(2000){Koubsk{\'y}}, {Harmanec}, {Hubert},
  {Floquet}, {Kub{\'a}t}, {Ballereau}, {Chauville}, {Bo{\v z}i{\'c}},
  {Holmgren}, {Yang}, {Cao}, {Eenens}, {Huang}, \& {Percy}}]{koubsky2000}
{Koubsk{\'y}}, P., {Harmanec}, P., {Hubert}, A.~M., {et~al.} 2000, \aap, 356,
  913

\bibitem[{{Kraus}(2016)}]{kraus2016}
{Kraus}, M. 2016, Boletin de la Asociacion Argentina de Astronomia La Plata
  Argentina, 58, 70

\bibitem[{{Maravelias} {et~al.}(2017){Maravelias}, {Kraus}, {Cidale}, {Arias},
  {Aret}, \& {Borges Fernandes}}]{maravelias2017}
{Maravelias}, G., {Kraus}, M., {Cidale}, L., {et~al.} 2017, in IAU Symposium,
  Vol. 329, The Lives and Death-Throes of Massive Stars, ed. J.~J. {Eldridge},
  J.~C. {Bray}, L.~A.~S. {McClelland}, \& L.~{Xiao}, 421--421

\bibitem[{{Naze} \& {Motch}(2018)}]{naze2018}
{Naze}, Y. \& {Motch}, C. 2018, ArXiv e-prints 1809.03341

\bibitem[{{Neiner} {et~al.}(2011){Neiner}, {de Batz}, {Cochard}, {Floquet},
  {Mekkas}, \& {Desnoux}}]{neiner2011}
{Neiner}, C., {de Batz}, B., {Cochard}, F., {et~al.} 2011, \aj, 142, 149

\bibitem[{{Neiner} \& {Mathis}(2014)}]{neiner2014}
{Neiner}, C. \& {Mathis}, S. 2014, in IAU Symposium, Vol. 301, Precision
  Asteroseismology, ed. J.~A. {Guzik}, W.~J. {Chaplin}, G.~{Handler}, \&
  A.~{Pigulski}, 465--466

\bibitem[{{Owocki} {et~al.}(2014){Owocki}, {ud-Doula}, {Townsend}, {Petit},
  {Sundqvist}, \& {Cohen}}]{owocki2014}
{Owocki}, S.~P., {ud-Doula}, A., {Townsend}, R.~H.~D., {et~al.} 2014, in IAU
  Symposium, Vol. 302, Magnetic Fields throughout Stellar Evolution, ed.
  P.~{Petit}, M.~{Jardine}, \& H.~C. {Spruit}, 320--329

\bibitem[{{Rivinius} {et~al.}(2013){Rivinius}, {Carciofi}, \&
  {Martayan}}]{rivinius2013}
{Rivinius}, T., {Carciofi}, A.~C., \& {Martayan}, C. 2013, \aapr, 21, 69

\bibitem[{{Secchi}(1867)}]{secchi1867}
{Secchi}, A. 1867, Astronomische Nachrichten, 68, 63

\bibitem[{{Slettebak} {et~al.}(1992){Slettebak}, {Collins}, \&
  {Truax}}]{slettebak1992}
{Slettebak}, A., {Collins}, II, G.~W., \& {Truax}, R. 1992, \apjs, 81, 335

\bibitem[{{Smith} {et~al.}(2016){Smith}, {Lopes de Oliveira}, \&
  {Motch}}]{smith2016}
{Smith}, M.~A., {Lopes de Oliveira}, R., \& {Motch}, C. 2016, Advances in Space
  Research, 58, 782

\bibitem[{{Telting} {et~al.}(1994){Telting}, {Heemskerk}, {Henrichs}, \&
  {Savonije}}]{telting1994}
{Telting}, J.~H., {Heemskerk}, M.~H.~M., {Henrichs}, H.~F., \& {Savonije},
  G.~J. 1994, \aap, 288, 558

\end{thebibliography}

\end{document}